# Depth First Always On Routing Trace Algorithm

Anthony Kim, *Member, IEEE,* Sung Hyun Chen, *Member, IEEE,* Chen Zheng, *Member, IEEE*

**Abstract**
**In this paper, we discussed current limitation in the electronic-design-automotation (EDA) tool on tracing the always on routing. We developed an algorithm to efficiently track the secondary power routing and accurately estimate the routing quality using approximate voltage drop as the criteria. The fast check can identify potential hotspot issues without going through sign-off checks. It helps designers to capture issues at early stages and fix the issues with less design effort. We also discussed some limitations to our algorithm.**

## 1. Introduction

In modern digital circuit design, to achieve lower power for mobile applications, multi power domain technique is widely used [1]. As a result, almost every design block has certain always-on (AON) instances that require AON routing to their secondary power pins [2]. Usually those routings are critical and are usually routed at the first priority [3]. However, those nets may get rerouted which degrade the quality, or there might be new AON instances added into design at later stages which have limited routing resource and result in bad AON routing [4]. Those issues often are not captured until sign-off checks [5], and designers will need to spend lot of effort on engineer-change-order (ECO) [6] to fix those issues. It is always desired to have checks that capture those issues at early stages to save design effort [7].

One challenge of tracing the quality of AON routing is that the secondary power pins are usually connected to global power nets (e.g. VDD supply), and the tracing feature in the tool will flag all occurrence of these nets and designers are not able to differentiate them for each AON instance. To overcome these limitations, we develop a depth first search (DFS) algorithm [8] that will track the AON routing from and to each secondary power pin and evaluate their voltage drop as routing quality. In Section 2, we describe the problem definition in details; in Section 3, we discussed the tracing algorithm and compared between forward search and backward search; in Section 4, we discuss memorizing the weight for current and voltage drop evaluation; in Section 5, we show experimental results and conclude in Section 6.

## 2. Problem Statement

In a typical modern low power digital design, VDD towers over power switch columns are often the anchor tapping point for AON instance secondary power pin [9][10]. Usually a limit of fanout number is applied such that the current on the trunk can be kept under a threshold value [11]. The VDD supply starts from the anchor point on the power switch tower and drives the loads eventually. Figure 1 illustrates a typical AON routing layout with tapping point form power switch towers. The problem is formulated below: for all AON instances, find its current path from its secondary power pin to the VDD supply tapping point and estimate the total voltage drop along the path. It has two parts: (1) trace the path of the current flow; (2) estimate voltage drop. We will discuss the two parts separately in the following sections.

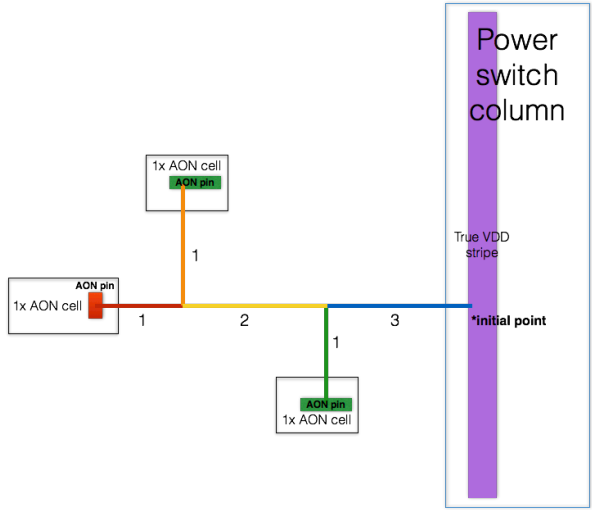

**Fig 1. AON routing topology**

## 3. Path Tracing

It is straightforward to apply DFS algorithm to trace the current flow paths. However, due to the non-deterministic property of DFS during its next candidate selection, this approach will incur redundant searches. In the worst case, it will result in $n*a$ searches, where $n$ is the number of fanout limit and $a$ is the number of AON instances. While ideally, the number of searches should equal to $a$ [12]. Since the trace of path from an AON instance power pin to tapping point is opposite in direction of the current flow, we refer this approach as backwards DFS search. Figure 2 shows a typical example of how redundant search could happen using backwards DFS [13].

To overcome this issue, we propose an approach that starts from the tapping point of VDD supply and follows the direction of current flow to trace all its driving loads. By doing this, it avoids the redundant searches and can also using stack to memorize the current weight during recursive calls. We referred to this as forward DFS. In Section 4, we discuss the current weight memorization in details.

## 4. Current Weight Memorization

As stated in previous sections, it is important to determine the correct weight for each AON instance current on each AON routing segment to accurately evaluate the approximate voltage drop along the AON routing path. In Figure 1, such current weight coefficient is shown along each current path segment. We follow the criteria below to determine the current weight: (1) the initial current weight of an AON instance is given by $D$, where $D$ is the cell size of the AON instance; (2) when multiple branches of leaves are merged into one trunk or intermediate branch, the current weight is accumulated; (3) worst case current is assumed for our analysis [14]. Such current weight calculation can be implemented using a stack associated with the recursive forward DFS call [15]. Every time when the child recursive call returns, the parent call will update the current weight. In this way, it also helps to identify large current density occurring on sensitive structures [16]. Figure 3 shows the procedure of how current weight coefficient is updated using stack memorization of forward DFS.

## 5. Experiments

We used two blocks from the open source design or1200_fcmp and or1200_genpc [17] as our benchmark testcases. The logical synthesis is done using Synopsis Design Compiler [18] and physical synthesis is done using Synopsis ICC2 [19]. Sign-off extraction is done using Synopsis Star-RC [20] and power integrity checks are run using Apache RedHawks [21]. Table 1 and 2 shows the comparison between our forward DFS algorithm and sign-off check results. It can be observed that our forward DFS AON routing check achieves high level of confidence compared to sign-off checks. Note that we could also apply technique such in [22] to speed up the checks. There are a few possibilities for the error of DFS algorithm that limit its accuracy: (1) the current drawn by each AON instance is only an estimation, so no accurate current is initialized; (2) the resistance along the path is computed using simple model, no complex modeling is considered such as resistivity dependency on proximity, so there is miscorrelation as compared to actual extraction; (3) the voltage drop at the tapping point is unknown, so the voltage drop estimation is only for the tapping point to its load; (4) metal fill impact [23] is ignored; (5) effects such as variation [24], temperature [25] and layout impact [26] is not considered.

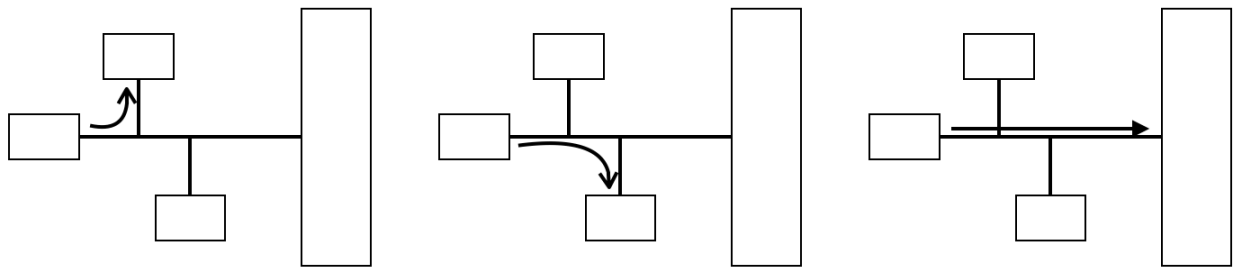

**Fig 2. Redundant Search by Backwords DFS**

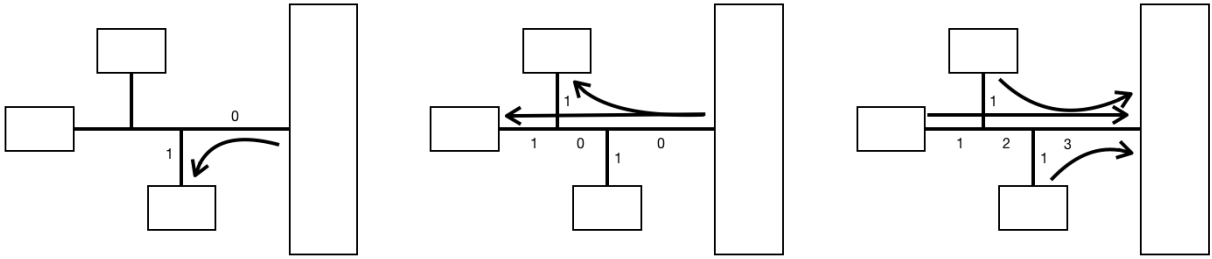

**Fig 3. Current Weight Update**

## 6. Conclusions

In this paper we show the limitation of current EDA tool to provide trace check for AON routings. We developed a forward DFS tracing algorithm that efficiently and accurately track AON routing for each AON instance power pin. The algorithm uses a stack memorization technique to update the current weight to evaluate the voltage drop. Experimental results show good match against sign-off checks. Several limitations in our proposed algorithm is discussed and can be improved further for more accuracy.

**Table 1. Comparison between sign-off and DFS checks**

|  | or1200_fcmp | or1200_genpc |
|---|---|---|
| #violation sign-off | 216 | 553 |
| #violation DFS | 255 | 612 |
| False positive | 69 | 86 |
| Missing positive | 36 | 74 |
| Accuracy | 83.3% | 86.6% |
| Runtime sign-off | 7h 35m | 12h 24m |
| Runtime DFS | 45m | 1h 4m |